\documentclass{aa}
\usepackage{psfig,natbib}

\def\feka{Fe K$\alpha$}
\def\chandra{{\it Chandra}} 
\def\xmm{{\it XMM-Newton}} 
\def\asca{{\it ASCA}} 
\def\hst{{\it HST}} 
 
\def\sax{{\it BeppoSAX}} 
 
\def\rosat{{\it ROSAT}}

\def\lum{erg s$^{-1}$}
\def\flux{erg cm$^{-2}$ s$^{-1}$}
\def\nh{cm$^{-2}$}
\def\arcsec{$^{\prime\prime}$}
\def\arcmin{$^{\prime}$}

\def\ngc{NGC~6251}

\def\ltsima{$\; \buildrel < \over \sim \;$}
\def\simlt{\lower.5ex\hbox{\ltsima}} 
\def\gtsima{$\; \buildrel > \over \sim \;$}
\def\simgt{\lower.5ex\hbox{\gtsima}} 

\begin{document}

\title{The XMM-Newton view of \object{NGC~6251}}

\author{M. Gliozzi\inst{1}  
\and  R.M. Sambruna\inst{1}
\and W.N. Brandt\inst{2}
\and R. Mushotzky\inst{3}
\and M. Eracleous\inst{2}} 
\offprints{mario@physics.gmu.edu} 
\institute{George Mason University, Dept. of Physics \&
Astronomy \& School of Computational Sciences, 4400 University Drive, MS 3F3, Fairfax, VA 22030
\and  The Pennsylvania State University, Department of
Astronomy \& Astrophysics, 525 Davey Lab, University Park, PA 16802
\and NASA Goddard Space Flight Center, Code 662, 
Greenbelt, MD 20771}

\date{Received: ; accepted: }

\abstract{We report on the nuclear X-ray properties of the radio
galaxy \object{NGC~6251} observed with 
\xmm. \ngc\ is a well-known radio galaxy 
with intermediate FRI/II radio properties.
It is optically classified as a Seyfert 2 and hosts a
supermassive black hole with mass $\sim 6\times10^8  M_{\odot}$.
The 0.4--10 keV EPIC pn continuum is best fitted by two thermal 
components ($kT \sim 0.5$ and 1.4 keV, respectively), plus a power law with 
photon index $\Gamma \sim 1.9$ absorbed by a
column density $N_{\rm H} \sim 5 \times 10^{20}$ \nh. 
We confirm the previous \asca\
detection of a strong iron line. The line, resolved in the EPIC pn spectrum,
is adequately fitted with a broad
($\sigma\sim 0.6$ keV) Gaussian at rest-frame energy 6.4 keV with
EW $\sim$ 220 eV.
We also detect, for the first time,
short-term, low-amplitude variability of the nuclear flux on a
timescale of a few ks.  The spectral properties argue in favor of the
presence of a standard accretion disk, ruling out the base of the jet
as the sole origin of the X-rays. The moderate X-ray luminosity and lack of 
strong intrinsic absorption suggest that \object{NGC~6251}
 is a ``pure'' type 2 AGN which lacks a broad-line region. 
}
\titlerunning{The XMM-Newton view of \object{NGC~6251}}
\authorrunning{M.~Gliozzi et al.}
\maketitle

\section{Introduction}

\object{NGC~6251} ($z$=0.024) is a giant elliptical galaxy hosting a
supermassive black hole with mass $M_{\rm BH}\sim 4-8 \times10^8
M_{\odot}$ (Ferrarese \& Ford 1999), as measured with \hst.  At radio
wavelengths, the source has been well studied. It shows the typical
morphology of a Fanaroff-Riley II (FR~II)
with a very narrow and long ($>4$\arcmin) jet extending N-W (e.g.,
Saunders et al. 1981), which emits at X-ray wavelengths 
(Birkinshaw \& Worrall 1993; Mack et al. 1997). 
However, based
on its radio power at 178 MHz, \ngc\ is classified as an FR~I
(e.g., Owen \& Laing 1989). Thus,
\ngc\ occupies an interesting niche in the study of radio galaxies as
an intermediate object (from the radio perspective) between low- and
high-power sources.

In the optical, \ngc\ has been classified as a Seyfert~2, based
on the presence of permitted and forbidden emission lines with FWHM
$\sim 600 {~\rm km~s^{-1}}$(e.g., Shuder \& Osterbrock 1981).
Previous X-ray studies of this galaxy with \rosat\ showed the presence
of an unresolved nuclear source embedded in diffuse thermal emission
associated with the galaxy's halo (Birkinshaw \& Worrall 1993). \asca\
observations showed that the nuclear spectrum could be described as a
moderately absorbed power law continuum and also suggested the
presence of a strong (EW$\sim 400$ eV), narrow line at $\sim$6.7 keV
(Sambruna et al. 1999; Turner et al. 1999).

Despite the intensive study of this source at all wavelengths, the
nature of the accretion process in \object{NGC~6251} is still a matter
of debate. Based on the radio-to-X-ray spectral energy distribution,
Ho (1999) suggested that an Advection-Dominated Accretion Flow (ADAF)
is present in the nucleus of \ngc. On the other hand, Ferrarese \&
Ford (1999) and Melia et al. (2002) favored a standard accretion
disk. Similarly, the origin of the X-ray emission from the nucleus is
uncertain: Turner et al. (1997) advocated a typical obscured Seyfert 2
spectrum based on the \asca\ data, while Hardcastle \& Worrall
(2002) argued in favor of an origin from the base of the relativistic
jet for the soft X-rays.  Further support for the
jet-dominated hypothesis was recently claimed by Chiaberge et al.
(2003) based on spectral energy distribution arguments. On the other hand,
Guainazzi et al. (2003) favor a scenario with two main spectral components:
a blazar-like spectrum dominating the high-flux state and a Seyfert-like
spectrum emerging during the low-flux state.

With its larger sensitivity in the 0.3--10 keV band, the EPIC camera
on-board \xmm\ is an ideal instrument for investigating the origin of the
X-rays and the nature of the accretion process in \object{NGC~6251}
through timing and spectral analyses. Motivated by the above
considerations, we observed \ngc\ with \xmm\ in AO1 for 50 ks. Here we
report the results from the analysis of the nuclear light curve and
spectrum. X-ray emission from the extended jet and halo will be
discussed elsewhere (Sambruna et al. 2003).  Throughout the paper we
use a Friedman cosmology with $H_0=75~{\rm
km~s^{-1}~Mpc^{-1}~and}~q_0=0.5$.

\section{Observations and data reduction}

We observed \object{NGC~6251} with \xmm\ on 2002 March 26 for a
duration of $\sim 41$ ks with the EPIC pn, and for $\sim 49$ ks with
EPIC MOS1 and MOS2. All of the EPIC cameras were operated in
full-frame mode with a medium filter for the MOS cameras and a thin
filter for the pn. The recorded events were screened to remove known
hot pixels and other data flagged as bad; only data with {\tt FLAG=0}
were used. The data were processed using the latest CCD gain
values, and only events corresponding to pattern 0--12 (singles,
doubles, triples, and quadruples) in the MOS cameras and 0--4 (singles
and doubles only, since the pn pixels are larger) in the pn camera
were accepted.  Arf and rmf files were created with the latest
available release of the \xmm\ Science Analysis Software (\verb+SAS+
5.4).  Investigation of the full--field light curves revealed the
presence of a period of background flaring at the end of the
observation.  These events were excluded, reducing the effective
total exposure time to $\sim 36$ ks for the EPIC pn and $\sim 43$ ks
for the MOS cameras. The RGS data of \object{NGC~6251} have a
signal-to-noise ratio ($S/N$) that is too low for a meaningful
analysis.  Background spectra and light curves were extracted from
source-free regions on the same chip as the source.  There are no
signs of pile-up in the pn or MOS cameras according to the {\tt SAS}
task {\tt epatplot}.  With an extraction radius of 35\arcsec\ the
detected count rates in the energy range 0.4--10 keV are
$(0.519\pm0.003){\rm~s^{-1}}$ for the MOS1,
$(0.533\pm0.004){\rm~s^{-1}}$ for the MOS2, and
$(1.810\pm0.007){\rm~s^{-1}}$ for the pn. For comparison with previous
broad-band X-ray satellites the count rates measured by the
SIS0 on-board ASCA and the MECS on-board \sax\ were 
$\sim 0.07 {\rm~s^{-1}}$ and $\sim 0.06{\rm~s^{-1}}$, respectively 
(Sambruna et al. 1999; Guainazzi et al. 2003).

Inspection of an archival \chandra\ observation of \object{NGC~6251} 
indicates that no
serendipitous sources are present within the EPIC extraction radius
and that the contribution from the resolved kpc jet is negligible.
However, no entirely reliable  information on the nuclear properties of
\object{NGC~6251} can be drawn from
\chandra\ data, due to the unfortunate location of the nucleus on
the CCDs' gaps and to the pile-up.

The EPIC spectra were rebinned such that each spectral bin contains at
least 100 counts for the pn and 40 counts for the MOS cameras
in order to apply $\chi^2$ minimization, and fitted
jointly using the {\tt XSPEC v.11.2} software package (Arnaud 1996). The
quoted errors on the derived best-fitting model parameters correspond
to a 90\% confidence level for one parameter of interest (i.e., a
$\Delta\chi^2=2.7$ criterion) unless otherwise stated.

\section{EPIC light curve of the nucleus} 

We first study the 0.3--10 keV light curve obtained with the EPIC pn, which
is the most sensitive instrument on-board \xmm,  using an extraction 
radius of 35\arcsec and  a time bin of
3000~s (changing the extraction radius or the time bin has a negligible
impact on our results). 
At energies below 0.8 keV no statistically significant variability is
detected according to a $\chi^2$ test. This might be partly due to a 
combination of dilution from the thermal emission and intrinsic absorption. 
However, the small intrinsic absorption and the relatively small
contribution of the thermal component in the low energy range (see $\S 4$)
suggest that the very-soft (0.4--0.7 keV) flux might be intrinsically
constant or, at most, variable at a very low level.
\begin{figure}
{\psfig{figure=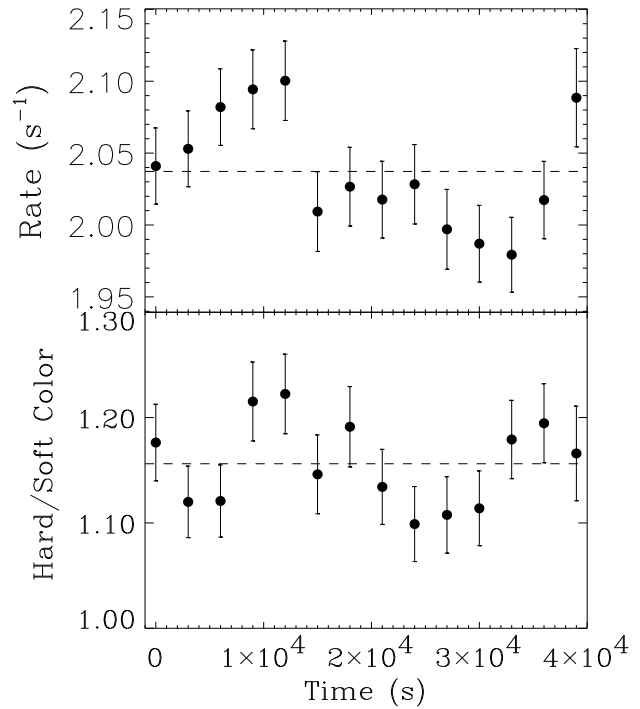,height=9.5cm,width=8.5cm,%
bbllx=30pt,bblly=5pt,bburx=425pt,bbury=455pt,angle=0,clip=}
\caption{EPIC pn+MOS1+MOS2 light curves of the background-subtracted
count rate in the 0.8--10 keV band (top panel) and of the X-ray color 
1.8--10 keV/0.8--1.5 keV (bottom panel).
The extraction radius is 35\arcsec; time bins are 3000 s. The
dashed lines indicate the average values.}
\label{figure:lc}}
\end{figure}
We therefore focus on the
0.8--10 keV energy band. The combination of pn, MOS1, and MOS2 light 
curves increases the statistical significance of the variability.
Figure~\ref{figure:lc} (top panel) shows the 0.8--10 keV
background-subtracted EPIC pn+MOS1+MOS2 light curve of \object{NGC~6251}. The
bottom panel shows a plot of the hardness ratio, defined as the
ratio of the 1.8--10 keV count rate to the 0.8--1.5 keV count rate, versus
time. Low-amplitude flux variability is present on timescales of a few ks. A
$\chi^2$ test yields a probability that the count rate is constant of
P$_{\chi^2}$=0.9\%.
The background count rate is as low as 5\% of the
average source count rate. The variability of the hardness ratio is not
significant (P$_{\chi^2}\sim$22\%).
\begin{table}[ht] 
\caption{Short-term X-ray variability of \object{NGC~6251}}
\begin{center}
\begin{tabular}{ccccc}
\hline
\noalign{\smallskip}
Energy band & $\langle r\rangle$ & $\chi^2_{\rm red}$ 
&  $P_{\chi^2}$ & $F_{\rm var}$ ${\rm ^a}$ 
\\ 
\noalign{\smallskip}       
(keV) & (${\rm s^{-1}}$) & (14 d.o.f.) & (\%) & (\%)\\
\noalign{\smallskip} 
\noalign{\smallskip}      
\hline
\hline
\noalign{\smallskip}
\noalign{\smallskip}
0.4--0.7 keV  & 0.49&  0.54 & 89.6 & $-$   \\
\noalign{\smallskip}
0.8--1 keV  & 0.34&  1.40 & 14.8 & $2.2\pm1.4$   \\
\noalign{\smallskip}
1.2--1.5 keV  & 0.36  & 1.48 & 11.7 & $2.1\pm1.3$   \\
\noalign{\smallskip}
1.8--3 keV   & 0.41   & 1.59 & 8.1 & $2.3\pm1.2$   \\
\noalign{\smallskip}
3--10 keV   & 0.39    & 1.66 & 6.2 & $2.8\pm1.3$   \\
\noalign{\smallskip}
\hline
\end{tabular}
\end{center}
${\rm ^a}$ The errors on $F_{\rm var}$ are calculated as in
Edelson et al. (2002) and should be considered conservative estimates
of the true uncertainty. 
\label{table:shortvar}
\end{table}

The study of energy-dependent variability is hampered by the presence
of photons emitted from the diffuse component that dilutes the
intrinsic variability of the source,
especially at low energies. Therefore, we
have 1) divided the 0.8--10 keV range into several sub-bands; 2)
checked with a $\chi^2$ test the presence of variability in each
sub-band; 3) merged contiguous variable sub-bands, creating four final
bands with roughly the same count rates. The energy-dependent
variability has been characterized by means of  the $\chi^2$ chance probability
and the fractional
variability parameter, $F_{\rm var}=(\sigma^2-\Delta^2)^{1/2}/\langle
r\rangle$, where $\sigma^2$ is the variance, $\langle r\rangle$ the
unweighted mean count rate, and $\Delta^2$ the mean square value of
the uncertainties associated with each individual count rate.  The
results are reported in Table~\ref{table:shortvar}. For comparison,
the 0.4-0.7 keV results are also reported.
No clear variability trend with energy is found.

\object{NGC~6251} has been observed by several X-ray satellites over the last
decade. Therefore, its long-term temporal behavior can be investigated.
A comparison of the 2--10 keV absorbed flux detected by ASCA in October 1994
(Sambruna et al. 1999), $F_{\rm ASCA}\sim 1.4\times10^{-12}$ \flux , with
values more recently measured with \sax\ 
(observation carried out in July 2001),
$F_{\rm SAX}\sim 4.7\times10^{-12}$ \flux, and with \xmm\ (see below)   
$F_{\rm XMM}\sim 4\times10^{-12}$ \flux, suggests an increase by a factor of
three over the last nine years. Although this conclusion might be 
weakened by
systematic errors related to the instrumental cross-calibration uncertainties,
the spectral uncertainties, and the different extraction areas, the substantial
flux difference measured argues in favor of genuine variability.
Observations with the same satellite
are needed to carefully investigate the long-term variability.

\section{The nuclear spectrum} 
\subsection{The continuum} 
\begin{figure}[htb]
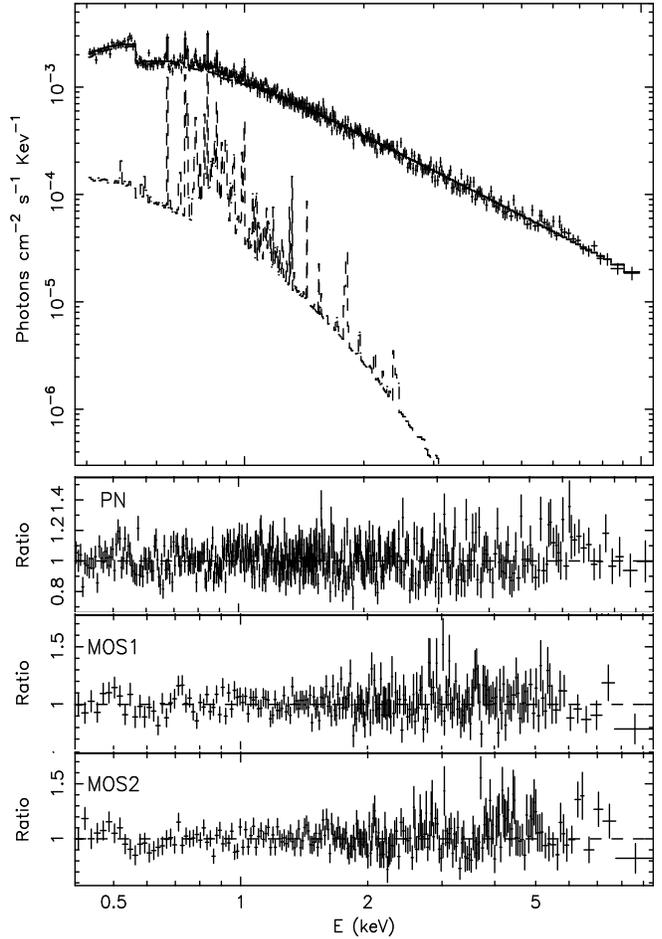

\psfig{figure=pn-2b.ps,height=6.2cm,width=8.7cm,%
bbllx=78pt,bblly=17pt,bburx=533pt,bbury=704pt,angle=-90,clip=}
\vskip 0.1 cm
\psfig{figure=pn-1b.ps,height=1.8cm,width=8.7cm,%
bbllx=384pt,bblly=17pt,bburx=515pt,bbury=704pt,angle=-90,clip=}
\psfig{figure=m1-2b.ps,height=1.8cm,width=8.7cm,%
bbllx=384pt,bblly=17pt,bburx=515pt,bbury=704pt,angle=-90,clip=}
\psfig{figure=m2-2b.ps,height=2.5cm,width=8.7cm,%
bbllx=384pt,bblly=17pt,bburx=570pt,bbury=704pt,angle=-90,clip=}
\caption{
EPIC pn unfolded spectrum of \object{NGC~6251} described by a thermal
component plus an absorbed power law. The lower panels show the
data-to-model ratio for the pn, MOS1, and MOS2 cameras, respectively.
The edge at 0.53 keV in the top panel is due to neutral oxygen in 
the absorbing material.
\label{figure:spec1}} 
\end{figure} 
Motivated by the controversial results of previous X-ray spectral
studies, we fitted the EPIC spectra in the 0.4--10 keV range, where
the instruments are best calibrated, with models of increasing
complexity.  To account for residual calibration problems between the
EPIC cameras, we first fitted separately the pn, MOS1, and MOS2
spectra. All the spectral models assume a column density, obscuring
all components, fixed at the Galactic value of $N_{\rm H}=5.65\times
10^{20}$ \nh\ (Dickey \& Lockman 1990). In some models an additional
intrinsic absorber at the redshift of \object{NGC~6251} is included. A
power-law model modified by Galactic absorption gives unacceptable
fits: $\chi^2/d.o.f.$ are 927.8/443, 401.2/232, 384.5/233, for the pn,
MOS1, and MOS2, respectively.  A first significant improvement
($\Delta\chi^2$\gtsima 80 for three additional parameters) is obtained
by adding a soft component modeled as emission by a
collisionally-ionized plasma (\verb+apec+ in \verb+XSPEC+; Smith et al.
2001).  However,
a fair representation of the EPIC spectra is only obtained with a
thermal component plus an absorbed power law. The column density of
cold gas toward the nucleus $N_{\rm H}\simeq 5 \times 10^{20}$ \nh,
although small, is statistically highly significant: $\Delta\chi^2\sim
70-100$ for one additional parameter.  We have investigated the
possibility that the intrinsic absorber is ionized (using the
\verb+absori+ in \verb+XSPEC+) or partially covering the nuclear
source (using the \verb+zpcfabs+ model in \verb+XSPEC+). However,
neither of these models resulted in a significant improvement of the fit.  
The
best-fit parameters for the model consisting of
a thermal component plus an absorbed power law, are reported in Table
2a (the abundances were frozen at the best-fit values).
Figure~\ref{figure:spec1} shows the unfolded spectrum of the EPIC
pn and the data-to-model ratio for the three EPIC cameras.
\begin{table*}[htb]
\caption{Spectral Model Parameters for \object{NGC~6251}.
Parameters without errors have been fixed at the best-fit values}
\begin{center}
\begin{tabular}{cccccccccc}
\hline
\hline
\noalign{\smallskip}
\noalign{\smallskip}
\multicolumn{10}{c}{{\bf a) EPIC continuum parameters}}\\
\noalign{\smallskip}
&\multicolumn{3}{c}{thermal} &\multicolumn{3}{c}{absorbed PL}&&& \\ 
\noalign{\smallskip}
\hline
\noalign{\smallskip}
Camera &  $kT$  & $Z$ & norm$_{\rm apec}$ &  $N_{\rm H}$ & 
$\Gamma$ &  norm$_{\rm pow}$ & $\chi^2/d.o.f.$ & Count rate &\\
 &keV   & $Z_\odot$ & $10^{-4}{\rm cm^{-5}}$ & $\rm 10^{20} cm^{-2}$ &
 & $10^{-3}~F_{\rm 1keV}^i$ & 
& ${\rm s^{-1}}$ &\\  
\noalign{\smallskip}
\hline
\noalign{\smallskip}
MOS1 & $0.59^{+0.09}_{-0.08}$ & 0.19 & $3.03^{+0.91}_{-0.82}$ &
 $6.5^{+1.4}_{-0.9}$ & $1.86^{+0.04}_{-0.04}$ & $1.19^{+0.05}_{-0.04}$&
 243.8/229 & $0.519\pm0.003$ &\\
\noalign{\smallskip}
\hline
\noalign{\smallskip}
MOS2 & $0.52^{+0.12}_{-0.11}$ & 0.18 & $3.03^{+1.07}_{-1.45}$ &
 $7.3^{+0.7}_{-1.8}$ & $1.92^{+0.03}_{-0.05}$ & $1.31^{+0.03}_{-0.06}$&
 236.0/230 & $0.533\pm0.004$ &\\
\noalign{\smallskip}
\hline
\noalign{\smallskip}
PN & $0.60^{+0.06}_{-0.05}$ & 0.18 & $3.30^{+0.76}_{-0.48}$ &
 $5.0^{+0.5}_{-0.7}$ & $1.91^{+0.03}_{-0.02}$ & $1.38^{+0.02}_{-0.05}$&
 548.4/440 & $1.810\pm0.007$ &\\
\noalign{\smallskip}
\noalign{\smallskip}
\hline
\hline
\noalign{\smallskip}
\noalign{\smallskip}
\multicolumn{10}{c}{{\bf b) Best-fitting continuum: EPIC pn}}\\
\noalign{\smallskip}
\multicolumn{3}{c}{thermal$_1$} &\multicolumn{3}{c}{thermal$_2$} &\multicolumn{3}{c}{absorbed PL}& \\ 
\noalign{\smallskip}
\hline
\noalign{\smallskip}
$kT_1$  & $Z_1$ & norm$_{\rm apec_1}$ &
$kT_2$  & $Z_2$ & norm$_{\rm apec_2}$ & $N_{\rm H}$ & 
$\Gamma$ &  norm$_{\rm pow}$ & $\chi^2/d.o.f.$ \\
keV   & $Z_\odot$ & $10^{-4}{\rm cm^{-5}}$ & 
keV   & $Z_\odot$ & $10^{-4}{\rm cm^{-5}}$ & $\rm 10^{20} cm^{-2}$ &
 & $10^{-3}~F_{\rm 1keV}^i$ &  ${\rm s^{-1}}$\\  
\noalign{\smallskip}
\hline
\noalign{\smallskip}
$1.36^{+0.34}_{-0.20}$ & 0.31 & $2.25^{+1.37}_{-1.05}$ &
$0.54^{+0.06}_{-0.08}$ & 0.17 & $3.23^{+0.72}_{-0.77}$ &
$5.1^{+1.1}_{-0.9}$ & $1.91^{+0.08}_{-0.05}$ & $1.91^{+0.08}_{-0.05}$&
520.3/435 \\
\noalign{\smallskip}
\noalign{\smallskip}
\hline
\hline
\noalign{\smallskip}
\noalign{\smallskip}
\multicolumn{10}{c}{{\bf c) Line parameters: EPIC pn}}\\
\noalign{\smallskip}
\hline
\noalign{\smallskip}
&&& E  & $\sigma$ & flux  &  EW  \\
&&&keV & keV & $ 10^{-6}~{\rm s^{-1}~cm^{-2}} $ & eV \\
\noalign{\smallskip}
\hline
\noalign{\smallskip}
&&&$6.42^{+0.52}_{-0.41}$ & $0.58^{+0.93}_{-0.24}$
& $8.5^{+8.3}_{-3.8}$  & $223^{+219}_{-99}$  \\
\noalign{\smallskip}
\hline
\end{tabular}
\end{center}
$^i {\rm ~photon~keV^{-1}~cm^{-2}~s^{-1}}$.
\label{table:spec}
\end{table*}
A close inspection of residuals shows a more complex pattern in the pn data.
Indeed the pn fit is further improved by adding a second thermal
component. This hotter thermal component is consistent with the one detected
in the circumnuclear gas around \object{NGC~6251} (Sambruna et al. 2003).
On the other hand, the fits to the MOS spectra are not improved significantly
because of their lower $S/N$.
We also tried an alternative model for the pn continuum with two power 
laws plus a thermal component. However, the resulting spectral fit, 
($kT=0.59_{-0.05}^{+0.06}$ keV, $Z=0.18~Z_\odot$, 
$\Gamma_1=2.68_{-0.33}^{+0.54}$, $\Gamma_2=1.63_{-0.03}^{+0.02}$) is 
statistically worse than the one obtained with two thermal models and one
absorbed power law ($\Delta\chi^2=+9.6$ for the same d.o.f.).

\begin{figure}[htb]
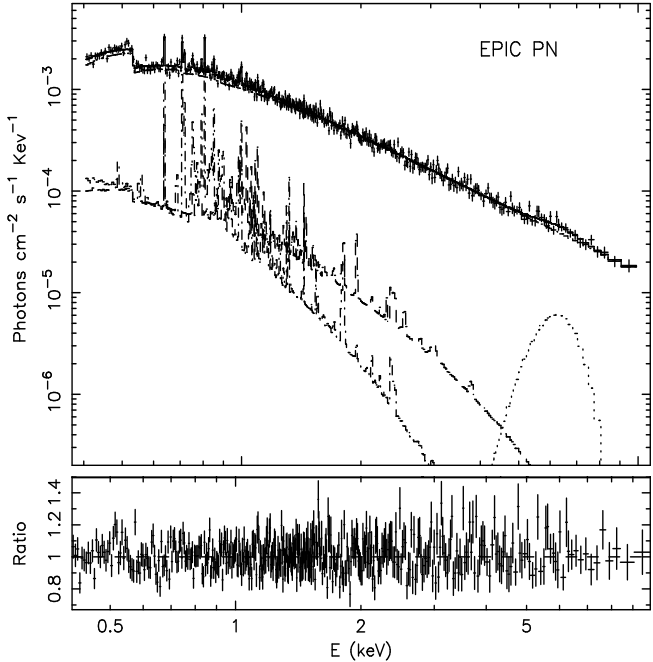

\psfig{figure=2apecB.ps,height=6.2cm,width=8.7cm,%
bbllx=78pt,bblly=17pt,bburx=533pt,bbury=704pt,angle=-90,clip=}
\vskip 0.1 cm
\psfig{figure=2apecA.ps,height=2.5cm,width=8.7cm,%
bbllx=384pt,bblly=17pt,bburx=570pt,bbury=704pt,angle=-90,clip=}
\caption{
EPIC pn unfolded spectrum of \object{NGC~6251} and data-to-model
ratio. The spectral model is wabs(apec+apec+zwabs(powerlaw+zgauss)).
\label{figure:spec2}} 
\end{figure}

In addition, the residuals of the pn and MOS2 data show a
clear excess in the 6--7 keV range, suggesting the presence of a line 
(see $\S 4.2$). 
The best-fitting continuum parameters are
summarized in Table 2b and the unfolded spectrum plus data-to-model ratio
is shown in Figure~\ref{figure:spec2}.
 
The total absorbed fluxes from 0.4--2 and 2--10 keV are $\sim 2.6 \times
10^{-12}$ \flux\ and $\sim 4.0 \times 10^{-12}$ \flux, respectively,
corresponding to intrinsic (absorption-corrected) luminosities of
L$_{0.4-2~\rm keV} \sim 4.4 \times 10^{42}$ \lum\ and L$_{2-10~\rm
keV} \sim 4.8 \times 10^{42}$ \lum. The
contribution of the thermal component is $\sim10$\% from
0.4--2~keV and negligible above 2~keV. This leads to an X-ray
luminosity associated with the power-law component of 
$L_{\rm0.4-10~keV}^{\rm PL}\sim 8.8 \times 10^{42}$ \lum. 
This value can be further increased up to $1.7\times
10^{43}$ \lum\ by extending the energy range up to 100 keV. In fact,
analysis of a \sax\ observation of \object{NGC~6251}
shows that the PDS detects the source up to 
energies higher than 100 keV (Guainazzi et al. 2003).

\subsection{The Fe line}
\begin{figure}
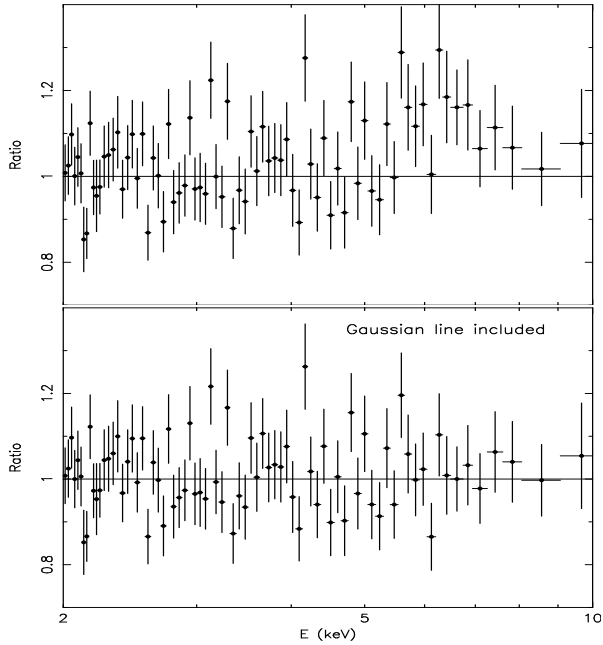

\psfig{figure=rat1-NEW.ps,height=4.cm,width=8cm,%
bbllx=81pt,bblly=20pt,bburx=532pt,bbury=710pt,angle=-90,clip=}
\psfig{figure=rat1b-NEW.ps,height=4.5cm,width=8cm,%
bbllx=81pt,bblly=20pt,bburx=590pt,bbury=710pt,angle=-90,clip=}
\caption{Top panel: EPIC pn data-to-model ratio from 2--10 keV of the best-fit
continuum model for the EPIC pn data. Bottom panel: same as above but with
a Gaussian line included. Energies are in the observer's frame.
\label{figure:res1}} 
\end{figure} 
To investigate the possible presence of an \feka\ line, we used only
pn data. In the MOS cameras, the rapid decrease of quantum efficiency with
increasing energy reduces the $S/N$ above 6 keV, hampering
a thorough analysis of the \feka\ range. 
To reduce the influence of the complex continuum modeling on the 
line detection and properties, we restrict the energy range to 2--10 keV.
The 5--7.5 keV band
was initially excluded to allow a better determination of the photon
index. A simple power law model produces a satisfactory fit to the
the 2--5 keV and 7.5--10 keV windows of the pn spectrum,
with  $\chi^2=134.1$
for 123 degrees of freedom. Including the 5--7.5 keV energy range, the
fit is still formally acceptable ($\chi^2=173.1$, 158 d.o.f). However,
residuals around 6--7 keV (Fig.~\ref{figure:res1} top panel)
suggest the presence of a broad line. Indeed, adding a Gaussian line
improves the fit significantly ($\Delta\chi^2$=14 for 3 additional
parameters), as can be seen from the residuals in the bottom panel
of Fig~\ref{figure:res1}. 
For testing purposes, we applied the same procedure to MOS2 data. 
A marginally significant  ($\Delta\chi^2=7.1$ for three additional parameters)
line at $\sim6.4$ keV is detected, with parameters consistent within the
errors with the ones inferred using pn data.  Using the entire
0.4-10 keV range and the best-fitting model, the statistical significance of 
the line detection is further enhanced: $\Delta\chi^2\simeq15.5$ for both pn 
and MOS2.

We also tried more complex fits to the broad line using specific disk
models in \verb+XSPEC+, such as {\tt diskline} and {\tt laor}. 
The former model parameterizes the expected line profile from a 
disk around a Schwarzschild black hole (Fabian et al. 1989), whereas
the latter describes the line
profile in the case of a rotating Kerr black hole (Laor 1991).
In both cases, the fitted inner radius of the disk is located
within a few Schwarzschild radii from
the black hole; however, due to the limited $S/N$, the line
parameters are poorly constrained.

In summary, we find evidence for a broad line at rest-frame energy
$\sim6.4$ keV with EW$\simeq220$ eV and width FWHM$\simeq64,000$ km ${\rm s^{-1}}$,
which we interpret as the fluorescent \feka\ line commonly
observed in Seyfert galaxies. The line parameters with the 90\% confidence
errors are given in Table 2c.

\section{Discussion and Conclusions}
A detailed discussion of the origin of the X-rays from weak radio galaxies
has been reported by Gliozzi et al. (2003) for the specific case of 
\object{NGC~4261}. For that object, a jet origin, claimed on the basis of a
tight correlation observed between radio and X-ray core luminosities,
was disputed on the basis of variability 
properties and energetic considerations. On the other hand, 
for \object{NGC~6251} rapid variability is detected but 
no clear flux or spectral variability trend is found, 
and the investigation of the origin of the X-rays must rely upon the 
time-averaged spectral results. 

The strongest argument against a jet origin  comes from the detection of a
prominent broad \feka\ line.  The large EW is consistent with the
average value found in Seyfert~1 galaxies (e.g., Nandra et al. 1997)
and with the upper limits found in brighter radio-loud AGN
(e.g., Sambruna et al. 1999), and is much larger than typical values found
in broad-line radio galaxies observed with RXTE (e.g., Eracleous et
al. 2000), arguing against a significant contribution from the jet.  A
Doppler-broadened line with FWHM$\simeq64,000$ km ${\rm s^{-1}}$ can
only be produced in the inner region of an accretion flow by
fluorescence (although a deeper X-ray exposure is necessary to
constrain the inner radius of the line-emitting region).
Contributions from the BLR are excluded here because the source lacks
optical broad lines  (e.g., Shuder \& Osterbrock
1981). Alternative scenarios invoking a jet with a wide opening angle
illuminating the putative obscuring torus can only produce narrow
emission lines (Wo\'zniak et al. 1998). 

In addition, the well
constrained photon index, $\Gamma\simeq1.9$, agrees very
well with the average value found in Seyfert 1 galaxies (e.g., Nandra
et al. 1997); however, it is not inconsistent with photon indices
derived for X-ray jets of FR~I radio galaxies, even though the latter
are generally poorly constrained (e.g., Hardcastle et al. 2002).

The small absorbing column is in good agreement with the low
visual extinction $A_{\rm V}\sim0.6$ inferred for the dusty disk
(Ferrarese \& Ford 1999), assuming the Galactic gas-to-dust ratio. The
lack of strong intrinsic absorption in \object{NGC~6251} argues against
obscuration by a canonical molecular torus (see, e.g., Chiaberge et
al. 2002).  The issue of intrinsic absorption in FR~I galaxies will be
addressed in a forthcoming paper (Donato et al. 2003, in prep).

Assuming that (most of) the X-rays are not associated with the jet, we
investigate the nature of the accretion flow. The main question is
whether an ADAF or a standard disk model provides a better description
of the accretion process in \object{NGC~6251}. To answer this
question, we first note that the ADAF interpretation proposed by
Ho (1999) was mainly prompted by the low value inferred for the ratio
$L_{\rm bol}/L_{\rm Edd}$, where $L_{\rm bol}=8.2\times 10^{42}$ \lum\
was obtained by integrating the spectral energy distribution from
radio to X-rays. However, because of the paucity of data points, the
estimate of $L_{\rm bol}$ is highly uncertain. Indeed, Ferrarese \&
Ford (1999) and Melia et al. (2002) independently derive a value of
$L_{\rm bol}$ larger by a factor of a few with respect to the value
derived by Ho.  Direct evidence that the value derived by Ho
underestimates $L_{\rm bol}$ comes from the value of X-ray luminosity
L$_{0.4-10~\rm keV}^{\rm PL}\sim 8.8\times 10^{42}$ \lum\ associated
with the power-law component. Therefore, assuming that $L_{\rm
0.4-10~keV}^{\rm PL}$ represents $\sim$30\% of $L_{\rm bol}$
 (see, e.g., Elvis et al. 1994), we
obtain a more reasonable but still conservative estimate of $L_{\rm
bol}$ of the order of $3\times 10^{43}$ \lum, consistent with the
estimate given by Ferrarese \& Ford (1999). However, the modest
value of the ratio $L_{\rm bol}/L_{\rm Edd}$ derived from this assumption
does not allow us to discriminate between the competing accretion scenarios.

An independent argument favoring a standard accretion disk is the 
detection of a broad \feka\ line. 
As already mentioned, the most natural explanation for the
broad line detected at $\sim 6.4$ keV is that it is a fluorescent
\feka\ line produced within the inner gravitational radii of a optically thick
accretion disk. 

Direct evidence against radiatively inefficient solutions comes
from the estimate of the radiative efficiency $\eta$, which can be
derived comparing the accretion luminosity $L_{\rm accr}=\eta \dot M_{\rm accr}
c^2$ to $L_{\rm bol}$. A rough estimate of the accretion rate is given by the
Bondi value, $\dot M_{\rm Bondi}=4\pi R_{\rm A}^2 \rho_{\rm A} c_{\rm s}$
where $R_{\rm A}\simeq GM/c_{\rm s}^2$ is the accretion radius, 
$c_{\rm s}$ the sound speed, and $\rho_{\rm A}$ the density 
at the accretion radius. Since the soft thermal component ($kT\sim
0.55$ keV) is spatially unresolved  with \xmm\ (Sambruna et al. 2003), it is
appropriate to choose the instrument resolution, 10\arcsec, as the angular size
of the thermal component. With this assumption, and considering 
normal cosmic composition ($n_{\rm e}/n_{\rm p}=1.18$), the emission measure
given by the normalization of the soft thermal component,
$ [10^{-14}/(4\pi(D_{\rm A}(1+z))^2)]\int n_e n_H dV\simeq 3\times 10^{-4}
{\rm~cm^{-5}}$ (where $D_{\rm A}$ is the angular size distance to the source 
in cm,and $n_e$ and $n_H$ are the electron and H densities in ${\rm cm^{-3}}$)
yields a proton density of $n_{\rm p}\simeq 0.06 {\rm~cm^{-3}}$, 
which translates into $\dot M_{\rm Bondi}\simeq 8\times
10^{-3}{\rm ~M_\odot~yr^{-1}}$.
A direct comparison
of this value with the estimated $L_{\rm bol}$ leads to a radiative efficiency
of the order of $\eta\sim0.06$, fully consistent with the standard
disk scenario.
This conclusion is supported by the striking difference observed
between \object{NGC~6251}, that radiates at the Bondi limit, and low luminosity
AGN observed with Chandra (Loewenstein et al. 2001)  which show 
$L_{\rm bol}\ll L_{\rm Bondi}$ and are interpreted in the ADAF framework.

A final important issue to be addressed is the nature of the Seyfert nucleus
in \object{NGC~6251}. Even though historically this source has been 
classified as a type 2 AGN,
its X-ray properties as seen with \xmm\ seem to be more typical
for a type 1 AGN.
Interestingly, we note that the value inferred for $L_{\rm bol}/L_{\rm Edd}$
translates into $\dot M/ \dot M_{\rm Edd}$ in the range where
recent models (e.g., Nicastro 2000, Laor 2003) predict that a broad-line 
region (BLR) cannot exist. The absence of broad optical lines and the 
low value inferred for the cold absorption from the X-ray spectral analysis
support the hypothesis that \object{NGC~6251} is 
a ``pure'' type 2 AGN, i.e., without BLR.
This result is consistent with the findings of Steffen et al. (2003), who, 
studying sources of the X-ray background, find that the fraction of objects 
with broad lines
drops sharply below $L_{\rm X}\sim 10^{44}{~\rm erg~s^{-1}}$.
An alternative interpretation might be that the
source is Compton-thick and the observed X-rays are only the scattered 
component. A possible way to test this hypothesis is based on the location of
\object{NGC~6251} in the Fe EW -- L$_{\rm X}$/L$_{\rm O III}$ plane 
(Bassani et al. 1999). However, such analysis does not provide any firm 
conclusion,
since \object{NGC~6251} is located half way between the Compton thick
region and the Seyfert 1 region. 
The rapid X-ray variability, the rather weak iron line 
for a type 2 AGN, and the
\sax\ results (Guainazzi et al. 2003)
seem to argue against the Compton-thick hypothesis.
 However, deep exposures at hard X-rays
are necessary to discriminate between the two competing scenarios.

A recent cross-correlation study of the FIRST and 2DF catalogs 
(Magliocchetti et al.
2002) finds that the majority of radio sources do not show any optical 
emission lines, while a minority have spectra similar to \object{NGC~6251}. 
If \object{NGC~6251} with its Seyfert 1-like nuclear properties
is representative of the whole group,
it follows that there is a significant fraction of sources which do not fit
the standard view of AGN. Detailed multiwavelength studies of their nuclear 
properties are needed.

\begin{acknowledgements} 
We thank the anonymous referee for the useful comments and suggestions
that improved the paper.
Financial support from NASA LTSA grants NAG5-10708 (MG, RMS),
NAG5-13035 (WNB), and NAG5-10817 (ME) is gratefully acknowledged. Funds
were also provided by NASA grant NAG5-10243 (MG, RMS) and by the Clare
Boothe Luce Program of the Henry Luce Foundation (RMS).
\end{acknowledgements}

\end{document}